\documentclass[twocolumn,showpacs,amsmath,amssymb,prb,superscriptaddress]{revtex4}

\usepackage{graphicx}
\usepackage{dcolumn}
\usepackage{bm}
\usepackage{color}
\usepackage{ulem}
\usepackage{amsfonts,amsthm}

\def\Vec#1{\mbox{\boldmath $#1$}}

\begin{document}

\newcommand{\rvec}{\Vec{r}}
\newcommand{\rvecp}{\Vec{r}^{\prime}}
\newcommand{\nablavec}{\Vec{\nabla}}
\newcommand{\qvec}{\Vec{q}}
\newcommand{\kvec}{\Vec{k}}

\newcommand{\bR}{\mbox{\boldmath $R$}}
\newcommand{\tr}[1]{\textcolor{red}{#1}}
\newcommand{\trs}[1]{\textcolor{red}{\sout{#1}}}
\newcommand{\tb}[1]{\textcolor{blue}{#1}}
\newcommand{\tbs}[1]{\textcolor{blue}{\sout{#1}}}
\newcommand{\Ha}{\mathcal{H}}
\newcommand{\mh}{\mathsf{h}}
\newcommand{\mA}{\mathsf{A}}
\newcommand{\mB}{\mathsf{B}}
\newcommand{\mC}{\mathsf{C}}
\newcommand{\mS}{\mathsf{S}}
\newcommand{\mU}{\mathsf{U}}
\newcommand{\mX}{\mathsf{X}}
\newcommand{\sP}{\mathcal{P}}
\newcommand{\sL}{\mathcal{L}}
\newcommand{\sO}{\mathcal{O}}
\newcommand{\la}{\langle}
\newcommand{\ra}{\rangle}
\newcommand{\ga}{\alpha}
\newcommand{\gb}{\beta}
\newcommand{\gc}{\gamma}
\newcommand{\gs}{\sigma}
\newcommand{\vk}{{\bm{k}}}
\newcommand{\vq}{{\bm{q}}}
\newcommand{\vR}{{\bm{R}}}
\newcommand{\vQ}{{\bm{Q}}}
\newcommand{\vga}{{\bm{\alpha}}}
\newcommand{\vgc}{{\bm{\gamma}}}
\arraycolsep=0.0em
\newcommand{\Ns}{N_{\text{s}}}


\title{Comparative study of hybrid functionals applied to structural and electronic properties of semiconductors and insulators
}

\author{Yu-ichiro Matsushita}
\email{matsushita@comas.t.u-tokyo.ac.jp}
\affiliation{%
Department of Applied Physics, The University of Tokyo, Hongo, Tokyo 113-8656, Japan}%

\author{Kazuma Nakamura}%
\affiliation{%
Department of Applied Physics, The University of Tokyo,
Hongo, Tokyo 113-8656, Japan}%

\author{Atsushi Oshiyama}%
\affiliation{%
Department of Applied Physics, The University of Tokyo, 
Hongo, Tokyo 113-8656, Japan}%
\affiliation{%
CREST, Japan Science and Technology Agency, Sanban-cho, Tokyo 102-0075, Japan}%

\date{\today}

\begin{abstract}
We present a systematic study that clarifies validity and limitation of current hybrid functionals in density functional theory for structural and electronic properties of various semiconductors and insulators. The three hybrid functionals, PBE0 by Perdew, Ernzerhof, and Becke, HSE by Heyd, Sucseria, and Ernzerhof, and a long-range corrected (LC) functional, are implemented in a well-established plane-wave-basis-set scheme combined with norm-conserving pseudopotentials, thus enabling us to assess applicability of each functional on equal footing to the properties of the materials. The materials we have examined in this paper range from covalent to ionic materials as well as a rare-gas solid whose energy gaps determined by experiments are in the range of 0.6 eV - 14.2 eV: i.e., Ge, Si, BaTiO$_3$, $\beta$-GaN, diamond, MgO, NaCl, LiCl, Kr, and LiF. We find that the calculated bulk moduli by the hybrid functionals show better agreement with the experiments than the generalized gradient approximation (GGA) provides, whereas the calculated lattice constants by the hybrid functionals and GGA show comparable accuracy. The calculated energy band gaps and the valence-band widths for the ten prototype materials show substantial improvement using the hybrid functional compared with GGA. In particular, it is found that the band gaps of the ionic materials as well as the rare-gas solid are well reproduced by the LC-hybrid functional, whereas those of covalent materials are well described by the HSE functional. We also examine exchange effects due to short-range and long-range components of the Coulomb interaction and propose an optimum recipe to the short-range and long-range separation in treating the exchange energy. 
\end{abstract}

\maketitle

\section{Introduction}\label{intro}

The local density approximation (LDA) (Ref.~\onlinecite{K-S}) in density functional theory (DFT) (Ref.~\onlinecite{H-K}) has shown fantastic performance in understanding and even predicting material properties\cite{LDA_appli} in spite of its relatively simple treatment of the exchange-correlation energy $E_{\rm XC} [n]$ as a functional of the electron density $n (\rvec)$; e.g., 
for many materials, lattice and elastic constants are generally reproduced. The deviations from experimental values are within less than 1-2 \% and several percent, respectively, in LDA. 
Yet LDA fails to describe some of properties including ground-state magnetic orderings even for bulk iron\cite{asada} and for some transition-metal oxides.\cite{oshiyama} It also tends to overestimate the bonding strength, leading to an absolute error of molecular atomization energies.\cite{perdew_Rev} 

Some of limitations of LDA are remedied by the generalized gradient approximation (GGA) in which the exchange-correlation energy is expressed in terms of not only the electron density but also its gradient. The molecular atomization energies are calculated with the error of several tenths of eV,\cite{kummel} and the ground state of the bulk iron is correctly predicted to be a ferromagnetic body-center phase.\cite{asada} The prevailed functional form of GGA (PBE) (Ref.~\onlinecite{pbe}) generally provides better accuracy for structural properties of variety of solids and activation energies in chemical reactions than LDA does.

The local (LDA) and semilocal (GGA) approximations are still insufficient to describe some of important properties, however. The ground states of strongly correlated materials are incorrectly predicted and the energy band gaps of most semiconductors and insulators are substantially underestimated. The meta-GGA scheme\cite{perdew99,voorhis,tao} extending the exchange-correlation functionals with including the kinetic energy density further improves the LDA and GGA results for molecular systems\cite{staroverov} but not succeed to remedy above failures in condensed matters. 

The failure of LDA and GGA is 
occasionally 
discussed in terms of the self-interaction error (SIE).\cite{SIC,cohen} An electron is under the electrostatic potential due to other electrons. Yet the expression of the electrostatic potential in the (semi)local approximations includes the spurious interaction with the electron itself. When we consider the Hartree-Fock (HF) exchange potential with Kohn-Sham orbitals, this spurious self-interaction is cancelled by a term in the exchange potential. In the (semi)local expression of the exchange potential, however, this cancellation is incomplete so that each electron is affected by the self-interaction. This SIE causes delocalization of the electron, predicting the incorrect fractional-charged ground state of, e.g., H$_2^+$ with large nucleus separation.\cite{cohen}

The SIE affects the band gaps substantially. The band gap $\Delta E_{\rm g}$ is formally defined as the ionization energy subtracted by the electron affinity so that $\Delta E_{\rm g} = E (N+1) + E(N-1) - 2 E(N)$ where $E(N)$ is the total energy of the $N$-electron system. In DFT with the exact exchange-correlation energy, the band gap is expressed as the difference between the highest occupied Kohn-Sham level $\varepsilon_{N+1} (N+1)$ of the $(N+1)$-electron system and its counterpart of the $N$-electron system $\varepsilon_N (N)$: i.e., $\Delta E_{\rm g} = \varepsilon_{N+1} (N+1) - \varepsilon_N (N)$.\cite{perdew82,perdew83,sham83,sham85} When we introduce a fractional electron system with $N+f$ electrons as a mixed state of real integer-electron systems, then the total energy $E(N+f)$ becomes linear for $0 < f < 1$ and shows discontinuity at the integer value $N$ for finite-gap systems. Using Janak theorem\cite{janak} which relates the Kohn-Sham level to the derivative of the total energy as $\varepsilon_{N+1} (N+f) = \partial E (N+f) / \partial f$, the linearity of $E (N+f)$ leads to the constant $\varepsilon_{N+1} (N+f)$ as a function of $f$. In the (semi)local approximations, however, the Kohn-Sham level $\varepsilon_{N+1} (N+f) $ [$\varepsilon_{N}(N-f)$] increases (decreases) with increasing $f$ due to the self-interaction, leading to the concave shape of $E(N+f)$. This may cause an underestimate of the energy gap.\cite{mori,cohen08} 

The HF approximation (HFA) is free from the self-interaction. Yet the calculated band gaps in HFA are substantially overestimated due to the lack of the correlation energy. An approach called the optimized effective potential\cite{talman} which is incorporated in DFT (Refs.~\onlinecite{ivanov,gorling,yang}) intending to remedy the issue is still in an immature stage in a view of applications to polyatomic systems. 

Hence the hybrid functionals combining LDA or GGA with HFA may be effective to break the limitation of the semilocal approximations. The hybrid approach has begun in empirical ways: The HF-exchange energy was mixed with the LDA exchange-correlation energy in the half and half way\cite{becke93A} and then three mixing parameters were introduced\cite{becke93B} to mix the LDA, GGA, and HFA energies; the latter scheme is called B3LYP and has widely been used to clarify thermochemical properties of molecules.\cite{B3LYP} A rationale for the hybrid functional is provided\cite{perdew96} in the light of the adiabatic-connection theorem,\cite{gunnarsson}
\begin{equation}
E_{\rm XC} \ [n] = \int_0^1 d\lambda \ E_{\rm XC,\lambda} \ ,
\label{adconn}
\end{equation}
where 
\begin{equation}
E_{\rm XC,\lambda} = \langle \Psi_{\lambda} | \hat{V}_{ee} | \Psi_{\lambda} \rangle 
                    - \frac{e^2}{2} \int d^3r \int d^3r^{\prime} 
                           \frac{n (\rvec) n(\rvecp)}{| \rvec - \rvecp | }
\label{EXCLam}
\end{equation}
is the energy of the exchange and correlation in a system, where the electron-electron interaction $\hat{V}_{ee} = (e^2/2) \sum_{ij} 1 / |\rvec_i - \rvec_j | $ is reduced by the factor $\lambda$ but the external potential $v_{\lambda} (r)$ is added to reproduce the electron density $n ( \rvec )$ of the real system ($\lambda=1$). Here $\Psi_{\lambda}$ is the ground-state many-body wave function. By assuming that $E_{\rm XC,\lambda}$ is the 4th polynomial of $\lambda$ with particular asymptotic forms for $\lambda=0$ and $\lambda=1$, Perdew, Ernzerhof, and Becke have proposed a parameter-free hybrid functional called PBE0\cite{perdew96} in which the HF and PBE exchange energies are mixed with the ratio of 1:3. Its applicability has been examined for molecular systems.\cite{adamo} 

Screening of the Coulomb potential is effective in polyatomic systems. Hence it may be effective to apply the non-local HF-exchange operator only to the short-range part of the Coulomb potential.\cite{bylander,heyd} This is conveniently done by introducing the error function splitting the Coulomb potential into short-range and long-range components. The hybrid functional which is constructed in this way from the PBE0 functional is proposed by Heyd, Sucseria, and Ernzerhof (HSE).\cite{heyd} This treatment reduces computational cost substantially and opens a possibility to apply the hybrid functionals to condensed matters. The structural properties as well as the band gaps of 
several 
solids have been calculated and significant improvements on semi-local functionals are achieved.\cite{paier,marsman,batista,fuchs,stroppa}

On the other hand, effects of the exchange interaction for the long-range component of the Coulomb potential are certainly important\cite{leininger,savin} in a view of reducing SIE. Hirao and his collaborators have proposed a long-range corrected (LC) functional in which the long-range component is treated by the HF exchange energy and the short-range component is by the LDA exchange.\cite{iikura} They have applied 
to various molecular systems and obtained relatively successful results.\cite{kamiya,tawada,chiba,sekino} Further application of the LC functional to molecular systems and its comparison with other functionals have been done and the applicability of the LC functional have been recognized.\cite{vydrov1,vydrov2} The LC functional has also been applied to structural properties and band gaps of 
several 
condensed matters and the results are compared with those obtained 
from other functionals.\cite{gerber} 

At the present stage, several hybrid functionals have been implemented in different packages and the assessment of the validity of each functional has been done mainly to molecular systems. Applicability of the PBE0 and HSE functionals to condensed matters have been examined.\cite{paier,marsman,stroppa} The structural and electronic properties such as lattice constants, bulk moduli, and also the band gaps of condensed matters are obtained 
only after careful examinations of various calculation parameters. Obviously, numerical precision should not be neglected.
In order to assess the validity of each hybrid functional, it is thus imperative to perform the computation in a single reliable calculation scheme. Furthermore, the split of the Coulomb potential into the short-range and long-range parts requires another parameter $\omega$ 
being the exponent of 
the error function. The $\omega$ dependence of the results should certainly be examined for better understanding and further improvements on the hybrid functionals. 

The aim of the present paper is to implement several important hybrid functionals in the well-established plane-wave-basis total-energy band-structure calculation code and examine validity and limitation of each functional.  A plane-wave code we adopt in this work is {\it Tokyo Ab initio Program Package} (TAPP).\cite{TAPP,sugino,yamauchi,kageshima} We have calculated lattice constants, bulk moduli, band gaps, and band widths of various semiconductors and insulators with the PBE0, HSE, and LC hybrid functionals as well as (semi)local GGA functional. Comparison of the obtained results unequivocally elucidates the validity and the limitation of the hybrid functionals.  

In section \ref{hybrid}, we briefly describe each of the hybrid functionals used in the present paper. Section \ref{computation} presents details of our computational scheme. The calculated results are 
shown in section \ref{result}, and our finding is summarized in section \ref{conclusion}.

\section{Hybrid exchange-correlation functionals}\label{hybrid}

In this section, we briefly describe the three hybrid functionals, PBE0, HSE, and LC, which we examine their applicability in this paper. 

\subsection{PBE0 functional}

Perdew, Burke, and Ernzerhof have proposed\cite{perdew96} a polynomial form for $E_{{\rm XC},\lambda}$ as 
\begin{equation}
  E_{\rm{XC},\lambda} 
= E_{\rm{XC},\lambda}^{{\rm DFT}} 
+ ( E_{\rm X}^{\rm{HF}} - E_{\rm X}^{\rm{DFT}}) ( 1-\lambda )^{n-1} ,
\label{eq:Exc_lambda} 
\end{equation}
where $ E_{{\rm X}}^{\rm{HF}}$ and $ E_{\rm X}^{\rm{DFT}} $ are the exchange energies obtained by HFA and a certain (semi)local approximation in DFT, respectively. Here $ E_{\rm{XC},\lambda=1}^{{\rm DFT}} $ is the energy of the exchange and correlation defined as Eq.~(\ref{EXCLam}) and obtained by the (semi)local approximation in DFT. This formula infers that the (semi)local approximation in DFT is a good approximation to $ E_{\rm{XC},\lambda}$ for $\lambda = 1$. 
When $\lambda = 0$, this formula equals to $ E_{{\rm x}}^{\rm{HF}}$ 
 since $ E_{\rm{XC},\lambda}^{\rm DFT} = E_{\rm{x}}^{\rm{DFT}} $ for $\lambda = 0$. 

From Eqs.~(\ref{adconn}) and (\ref{eq:Exc_lambda}), we obtain
\begin{equation}
  E_{\rm{XC}} 
= E_{\rm{XC}}^{\rm{DFT}} 
+ \frac{1}{n} ( E_{{\rm X}}^{\rm{HF}} - E_{\rm{X}}^{\rm{DFT}}). 
\label{PBE0_formula} 
\end{equation}
Relying on the fourth-order M\"{o}ller-Plesset perturbation theory applied for molecular systems, it is argued that $n=4$ is the best choice.\cite{perdew96} Using the PBE functional\cite{pbe} as the approximation in DFT in Eq.~(\ref{PBE0_formula}), the PBE0 hybrid functional is given by 
\begin{equation}
  E_{\rm{XC}} 
= E_{\rm{XC}}^{\rm{PBE}} 
+ \frac{1}{4} ( E_{{\rm X}}^{\rm{HF}} - E_{\rm{X}}^{\rm{PBE}}) , 
\label{PBE0_formula2} 
\end{equation}
leading to the mixing of 25 \% HF exchange and 75 \% PBE exchange. 

\subsection{HSE functional}

Heyd, Sucseria, and Ernzerhof have proposed\cite{heyd} a different hybrid functional in which the long-range part of the HF-exchange energy is treated by the semilocal approximation in DFT and the short-range part is calculated exactly. The actual procedure is conveniently done by splitting the Coulomb potential as 
\begin{equation}
\frac{1}{r} = 
\frac{{\rm erfc} (\omega r)}{r} + \frac{{\rm erf} (\omega r)}{r} ,
\label{split}
\end{equation}
and applying the first term only, i.e., the screened Coulomb potential, to the HF-exchange energy. 
The second term to the exchange energy is calculated with GGA. Adopting the mixing ratio in PBE0, then the HSE hybrid functional becomes 
\begin{equation}
E^{\rm HSE}_{\rm XC} = E_{\rm XC}^{\rm PBE} 
+ \frac{1}{4} ( E_{\rm X}^{\rm HF,SR} - E_{\rm X}^{\rm PBE,SR} ) . 
\label{HSE_formula}
\end{equation}
$E_{\rm X}^{\rm HF,SR}$ is the Fock-type double integral with the screened Coulomb potential. There is some complexity to divide the PBE exchange energy $ E_{\rm X}^{\rm PBE} $ into the short-range part $ E_{\rm X}^{\rm PBE,SR} $ and the long-range part $ E_{\rm X}^{\rm PBE,LR} $. 
The dividing procedure will be shown in subsection \ref{SRLR}.

Another ambiguous factor is the parameter $\omega$ in Eq.~(\ref{split}) which defines the short-range and the long-range parts of the Coulomb potential. The optimum value of $\omega$ = 0.15 $a_B^{-1}$ ($a_B$: Bohr radius) is proposed by examining the calculated results for molecular systems.\cite{heyd,heyd04,heyd04b,heyd06} Examining the $\omega$ dependence of the calculated results for condensed matters is one of our aims in this paper. 

\subsection{LC functional} 

The HSE functional partly removes SIE by incorporating the HF-exchange energy in the PBE functional. Yet the cancellation of the Hartree potential and the exchange potential is absent in the long-range part. This may cause erroneous description of, e.g., the Rydberg states in isolated polyatomic systems or properties of charge-transfer systems. To remedy this point, application of the long-range part of the Coulomb potential to the HF-exchange energy is necessary. The long-range corrected (LC) functional has been proposed based on this viewpoint,\cite{iikura} being expressed as 
\begin{equation}
E^{\rm LC}_{\rm XC} = E_{\rm XC}^{\rm DFT} 
+ ( E_{\rm X}^{\rm HF,LR} - E_{\rm X}^{\rm DFT,LR} ) ,
\label{LC_formula}
\end{equation}
with $E_{\rm X}^{\rm HF,LR}$ being the Fock-type double integral with the long-range part of the Coulomb potential [the second term of Eq.~(\ref{split})]. For the DFT part, several approximations including LDA,\cite{vydrov1,gerber}, PBE,\cite{tawada,vydrov1,vydrov2} and other GGA (Ref.~\onlinecite{iikura}) or meta-GGA (Ref.~\onlinecite{vydrov1}) forms are adopted in Eq.~(\ref{LC_formula}) and their validities are examined. It is argued that PBE combined in the LC hybrid scheme provides good accuracy for molecular properties.\cite{vydrov2} As in the HSE scheme, there is some complexity to extract the long-range component of the exchange energy of the DFT part, $ E_{\rm X}^{\rm DFT,LR} $. 

The parameter $\omega$ in Eq.~(\ref{split}) affects the results substantially. By examining the results for molecular systems, the values ranging $\omega$ = 0.25-0.5 are argued to be optimum.\cite{iikura,vydrov1} By applying LDA in the LC-hybrid scheme to structural properties of solids, the value $\omega$ = 0.5 is found to produce reasonable results.\cite{gerber} It is of interest to investigate the appropriate value of $\omega$ in the application of PBE in the LC-hybrid scheme to structural properties and band gaps of condensed matters. 

\section{Computational details}\label{computation}

In this section, we describe our implementation of the hybrid functionals in the plane-wave-basis-set total-energy band-structure calculation code, TAPP.\cite{TAPP,sugino,yamauchi,kageshima} Nuclei and core electrons are simulated by either norm-conserving\cite{troullier} or ultrasoft\cite{vanderbilt} pseudopotentials in the TAPP code. Non-locality of the HF-exchange potential generally increases computational cost tremendously in the application to condensed matters. We have circumvented this problem using the Fast-Fourier transform (FFT), as explained below. The long-range nature of the Coulomb potential leads to singularity of its Fourier transform at the origin. This causes difficulty in numerical integration over the Brillouin zone (BZ) to obtain the HF-exchange energy in the PBE0 and LC schemes. We here adopt a simple truncation scheme to overcome the problem. Finally we explain how to divide the exchange energy in the PBE functional to the short-range and long-range components. 

\subsection{Calculation of $E_{{\rm X}}^{{\rm HF}}$ by FFT}

The HF-exchange energy $E_{{\rm X}}^{{\rm HF}}$ in condensed matters (crystal) are written as 
\begin{equation}
    E_{\rm{X}}^{\rm{HF}} 
  = - \frac{1}{2} \sum_{n{\bf k}} \sum_{n'{\bf k'}} f_{n{\bf k}} f_{n'{\bf k'}}
      J_{n{\bf k}n'{\bf k'}} \ ,
\label{HFX_solid} 
\end{equation}
where  $f_{n{\bf k}}$ is the occupation number, and $J_{n{\bf k}n'{\bf k'}}$ is given by 
\begin{equation}
J_{n{\bf k}n'{\bf k'}}\!=\!\int\!\int\!d{\bf r}\!d{\bf r'}\! 
      \frac{\phi^*_{n {\bf k }} (\!{\bf r }\!)
            \phi^*_{n'{\bf k'}} (\!{\bf r'}\!)
            \phi_{  n {\bf k }} (\!{\bf r'}\!)
            \phi_{  n'{\bf k'}} (\!{\bf r }\!)}
            {|{\bf r}\!-\!{\bf r'}|} \ .
\label{Jij}
\end{equation}
Here $\phi_{n{\bf k}}({\bf r})$ is 
the Bloch-state orbital 
with the band index $n$ and the wavevector ${\bf k}$. The orbital $\phi_{n{\bf k}}({\bf r})$ is obtained by solving selfconsistently the Euler equation (Kohn-Sham equation) 
%
in which the exchange-correlation potential is given by the functional derivative of the hybrid exchange-correlation energy. 
The sum over ${\bf k}$ and $n$ are taken for the occupied states.

Our algorithm to compute the integral $J_{n{\bf k}n'{\bf k'}}$ in the plane-wave-basis code is as follows: Eq.~(\ref{Jij}) is written as
\begin{eqnarray}
\nonumber J_{n{\bf k}n'{\bf k'}}=\int d{\bf r} d{\bf r'} &&\frac{u_{n{\bf k}}^*({\bf r})u_{n'{\bf k'}}^*({\bf r'})u_{n{\bf k}}({\bf r'})u_{n'{\bf k'}}({\bf r})}{|{\bf r}-{\bf r'}|}\\
&&\times e^{-i({\bf k}-{\bf k'})\cdot{\bf r}} e^{i({\bf k}-{\bf k'})\cdot{\bf r'}},
\end{eqnarray}
with $u_{n{\bf k}}({\bf r})=\exp(-i{\bf k}\cdot {\bf r})\phi_{n{\bf k}}({\bf r})$. In the plane-wave-basis-set scheme, 
the reciprocal lattice vectors {\bf G} are used to represent the periodic wave function as $u_{n{\bf k}}({\bf r})=\sum_{\bf G}e^{i{\bf G}\cdot {\bf r}}u_{n\bf{k}}(\bf{G})$. We then obtain 
\begin{eqnarray}
\nonumber J_{n{\bf k}n'{\bf k}'}=&&\sum_{\bf G G' G'' }\frac{4\pi}{|{\bf k'-k+G'-G''}|^2} u_{n{\bf k}}({\bf G})^*u_{n'{\bf k}'}({\bf G}')^* \\ 
&\times& u_{n{\bf k}}({\bf G}'')u_{n'{\bf k}'}({\bf G+G'-G''}).
\end{eqnarray} 
When we compute Eq.~(\ref{Jij}) directly, the calculation costs of $J_{n{\bf k}n'{\bf k}'}$ and $E_{\rm{x}}^{\rm{HF}}$ are $O(N_{\bf G}^3)$ and $O(N_{\rm band}^2)\times O(N_{\bf k}^2)\times O(N_{\bf G}^3)$, respectively, where $N_{\bf G}$ is the total number of the reciprocal vectors,  $N_{\bf k}$ the total number of sampling $\bf k$ points in BZ, $N_{\rm band}$ the total number of the occupied bands. The number of $N_{\bf G}$ is much bigger than either $N_{\rm{band}}$ or $N_{\bf{k}}$. Hence the order $N_{\bf G}^3$ is computationally demanding.

We reduce this computational cost by using FFT. We first define the overlap density between states ($n{\bf k}$) and ($n'{\bf k'}$) as 
\begin{eqnarray}
  n_{n{\bf k}n'{\bf k'}}({\bf r})
= u^*_{n{\bf k}}({\bf r}) 
  u_{n'{\bf k'}}({\bf r}).
\end{eqnarray}
Using this quantity, we obtain 
\begin{eqnarray}
 \nonumber J_{n{\bf k}n'{\bf k'}}=\int d{\bf r}d{\bf r'} &&\frac{n_{n{\bf k}n'{\bf k'}}({\bf r})n_{n{\bf k}n'{\bf k'}}({\bf r})^*}{|{\bf r}-{\bf r}'|}\\
 &&\times e^{-i({\bf k}-{\bf k'})\cdot {\bf r}} e^{i({\bf k}-{\bf k}') \cdot {\bf r'}}.
\label{Jijk}
 \end{eqnarray}
 We note that, since $u_{n{\bf k}}({\bf r})$ has 
unit-cell periodicity, the overlap density also has the same periodicity. Using the Fourier transformation of the overlap density $n_{n{\bf k}n'{\bf k'}}({\bf r})$, Eq.~(\ref{Jijk}) is rewritten as 
\begin{eqnarray}
 J_{n\bf{k}n'\bf{k}'}=\sum_{\bf G}\frac{4\pi}{|\bf k - \bf k' - \bf G |^2}| n_{n{\bf k}n'{\bf k'}}({\bf G})|^2.
\label{Jijkl}
\end{eqnarray}
Since 
the $n_{n{\bf k}n'{\bf k'}}({\bf G})$ can be calculated outside the summation loop in Eq.~(\ref{Jijkl}), 
the total calculation cost for $J_{n\bf{k}n'\bf{k}'}$ is $O(N_{\bf G}+N_{\bf G}\ln N_{\bf G})$. The computational cost of $E_{\rm x}^{\rm HF}$ becomes $O(N_{\rm band}^2)\times O(N_{\bf k}^2)\times O(N_{\bf G}+N_{\bf G}\ln N_{\bf G})$. 
With considering the scaling as $N_{\bf G} \propto N_{\rm{atom}}$, $N_{\rm band} \propto N_{\rm{atom}}$, and  $N_{\rm{k}} \propto 1/N_{\rm{atom}}$ with $N_{\rm{atom}}$ the number of atoms in the unitcell, the computational cost above is proportional to 
$N_{\rm{atom}}+N_{\rm{atom}}\ln N_{\rm{atom}}$. 

\subsection{Treatment of divergence in Coulomb interaction}

The Fourier transform $ v (\qvec) $ of the Coulomb potential $v(\rvec)$ diverges at the long-wave length limit $ \qvec \rightarrow 0 $. Calculations of electrostatic energies thus require careful treatment and the well-known Ewald summation is the typical example. In calculating non-local exchange energies in condensed matters, more careful treatment is necessary. We need to perform the BZ integration in evaluation of the exchange energy in Eq.~(\ref{HFX_solid}). The integration is usually performed by the summation with weighting factors of the integrand at finite discrete $\kvec$ points, and thus encounter a difficulty to evaluate the integral accurately by picking up the singular behavior of the Fourier transform of the Coulomb potential. There are several ways to overcome the difficulty. One is the auxiliary-function approach: An auxiliary function which has the same singular behavior but is integrable is subtracted so that the summation can be done properly and the remaining term is obtained analytically.\cite{gygi,wenzien,carrier} An alternative way which we adopt in the present paper is simpler: We make the Coulomb potential truncated at $R_c$ and then examine the convergence by numerically increasing $R_c $.\cite{spencer} Namely, we replace Coulomb potential with a truncated potential,
\begin{equation}
\tilde{v}(\rvec) = 
\left\{ \begin{array}
{l}\frac{1}{| \rvec |}  \ \ \ \ \mathrm{if}\    \ | \rvec | \leq R_c \\ \\
0 \  \ \ \ \    \mathrm{otherwise}
\end{array}
\right.  \ . 
\end{equation}
with a cutoff radius $R_c$. What we need is of course converged quantities with $R_c \rightarrow \infty$. The truncated potential produces its non-divergent Fourier transform, 

\begin{equation}
\tilde{v} (\qvec) 
= \frac{4\pi}{| \qvec |^2} 
  \bigr[ 1 - \cos ( | \qvec | R_c ) \bigl]  \ .
\end{equation}

Convergence of required quantities with respect to $R_c$ is combined with the number of sampling $\kvec$ points in the BZ integration. The number $N_k$ of sampling $\kvec$ points required should increase with increasing $R_c$. We need to know the converged values with increasing both $R_c$ and $N_k$. The process to check this convergence in the two-parameter space can be done conveniently by introducing a certain relation between $R_c$ and $N_k$. We set up a relation, $N_k = 4 \pi R_c^3 / (3 \Omega_c )$, where $\Omega_c$ is the unit-cell volume, and examine the convergence of the exchange energy by increasing $N_k$ and equivalently $R_c$. We have found that the values of $R_c$ which are 4-6 times the dimension of the primitive unit cell are enough to assure the converged exchange energies in ten materials calculated in the present paper. 

\begin{figure*}
\begin{center}
\includegraphics[width=0.7\textwidth]{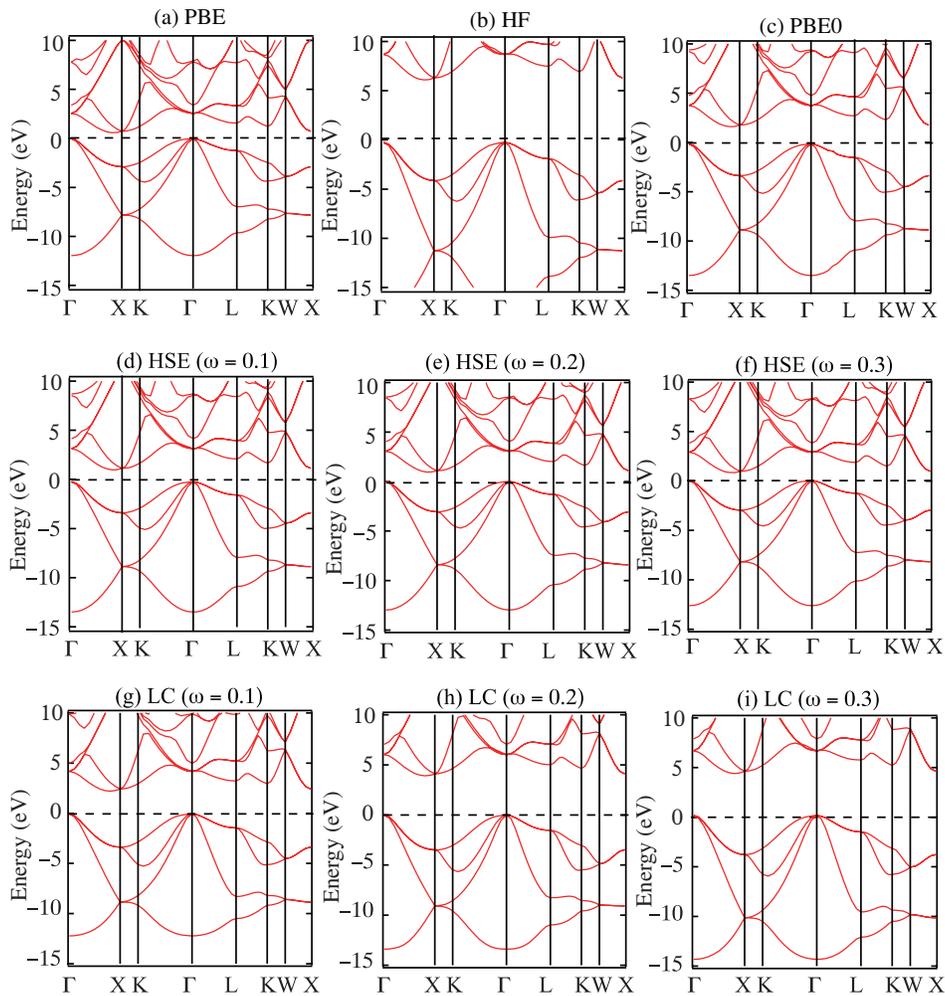}  
\end{center}
\caption{(Color online) Calculated energy bands of Si with PBE (a), HF (b), PBE0 (c), HSE (d)-(f), and LC (g)-(i) exchange-correlation functionals. The origin of the energy is set at the valence band top.}
\label{Si_band}
\end{figure*}

\subsection{Calculations of $E_{{\rm x}}^{{\rm PBE,SR}}$ and $E_{{\rm x}}^{{\rm BBE,LR}}$}\label{SRLR}

We next describe how to obtain the short-range and long-range parts in the PBE-exchange energy, following Ref.~\onlinecite{holemodel}. The original expression for the PBE-exchange energy is given by, 
\begin{eqnarray}
  E_{\rm{x}}^{\rm{PBE}} 
= \int d{\bf r} 
  \epsilon_{\rm{x}}^{\rm{unif}}[n({\bf r})] n({\bf r}) 
  F_{\rm{x}}^{\rm{PBE}}(s) \ ,
\label{TOT_PBE_X}
\end{eqnarray}
where $\epsilon_{\rm{x}}^{\rm{unif}}[n] $ is the exchange energy of the homogeneous electron gas with the electron density $n$ and $F_{{\rm x}}^{{\rm PBE}}(s)$ is an enhanced factor due to a density gradient $s\!=\!|\nabla n|/(2k_{\rm{F}} n)$ with the Fermi wavenumber $k_{\rm{F}}\!=\!(3\pi^2 n)^{1/3}$. The enhanced factor is written in an integral form
\begin{eqnarray}
  F_{\rm{x}}^{\rm{PBE}}(s) 
= -\frac{8}{9} \int_0^{\infty}dy y J_{{\rm x}}^{\rm{PBE}}(s,y),
\label{TOT_PBE_F}
\end{eqnarray}
where $y\!=\!k_{F}r$ is a dimensionless quantity and $J_{{\rm x}}^{{\rm PBE}}(s,y)$ describes an exchange-hole density at the distance $r$. Heyde, Scuseria, and Ernzerhof\cite{heyd} proposed an expression for the short-range PBE-exchange functional, where the original Coulomb interaction $1/r$ is modified to a screened form ${\rm erfc}(\omega r)/r$.
This modification leads to an enhanced factor somewhat different from the original expression of Eq.~(\ref{TOT_PBE_F}) as 
\begin{eqnarray}
F_{\rm{x}}^{\rm{PBE,SR}}\!(s,\omega) 
\!=\!-\frac{8}{9}\!\int_0^{\infty} dy y J_{{\rm x}}^{\rm{PBE}}(s,y) 
{\rm{erfc}} \Biggl(\!\frac{\omega y}{k_{\rm{F}}}\!\Biggr)\!. 
\label{SR_PBE_F} 
\end{eqnarray}
The short-range exchange energy is given with this enhanced factor as 
\begin{eqnarray}
  E_{\rm{x}}^{\rm{PBE,SR}}(\omega) 
\!=\!\int\!d{\bf r} 
  \epsilon_{\rm{x}}^{\rm{unif}} [n({\bf r})] 
  n({\bf r}) F_{{\rm x}}^{{\rm PBE,SR}}(s,\omega).
\label{SR_PBE_X} 
\end{eqnarray} 
The long-range PBE-exchange term is defined by subtracting the short-range part in Eq.~(\ref{SR_PBE_X}) from the original one in Eq.~(\ref{TOT_PBE_X}) as 
\begin{eqnarray}
  E_{{\rm x}}^{{\rm PBE,LR}}(\omega) 
= E_{{\rm x}}^{{\rm PBE}} 
- E_{{\rm x}}^{{\rm PBE,SR}}(\omega).
\end{eqnarray} 
Implementation details for these calculations can also be found in 
Ref.~\onlinecite{Heyde_Doctral_thesis}. 

\begin{figure*}
\begin{center}	
\includegraphics[width=0.7\textwidth]{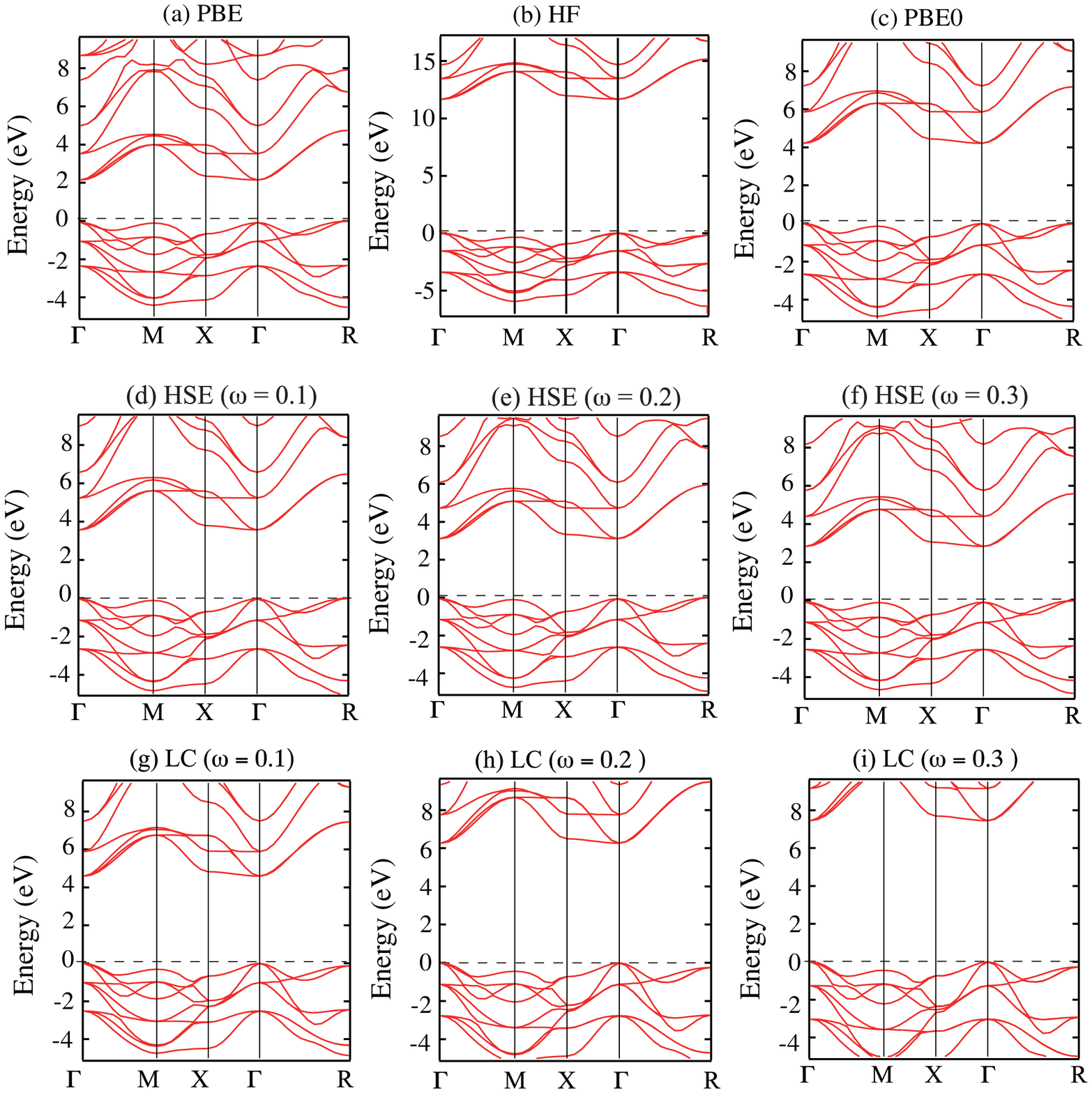}  
\end{center}
\caption{(Color online) Calculated band energy bands of BaTiO$_3$ with PBE (a), HF (b), PBE0 (c), HSE (d)-(f), and LC (g)-(i) exchange-correlation functionals. The origin of the energy is set at the valence band top. Note that the vertical scale in (b) is different from those in other panels.}
\label{BaTiO3_band}
\end{figure*}

\subsection{Calculation conditions} 

We generate norm-conserving pseudopotential to simulate nuclei and core electrons, following a recipe by Troullier and Matins.\cite{troullier} The core radius $r_c$ is an essential parameter to determine transferability of the generated pseudopotential. We have examined $r_c$ dependence of the calculated structural properties of benchmark materials and adopted the pseudopotentials generated with the following core radii in this paper: 0.85 {\AA} for Si 3$s$, and 1.16 {\AA} for Si 3$p$,  1.06 {\AA} for Ge 4$s$ and 4$p$, 1.06 {\AA} for Ga 4$s$ and 4$p$, and  1.48 {\AA} for Ga 4$d$,  0.64 {\AA} for N 2$s$ and 2$p$, 0.85 {\AA} for C 2$s$ and 2$p$, 0.79 {\AA} for O 2$s$ and 2$p$,  1.38 {\AA} for Na 2$s$ and 2$p$, 1.16 {\AA} for Cl 3$s$ and 3$p$, 0.95 {\AA} for Li 2$s$, 0.64 {\AA} for F 2$s$ and 2$p$, 1.59 {\AA} for Ba 5$s$, 5$p$, and 5$d$, 1.32 {\AA} for Mg 2$s$ and 2$p$, and 1.38 {\AA} for Ti 3$d$ and 4$s$, and 1.43{\AA} for Ti 4$p$, 1.48 {\AA} for Kr 4$s$ and 1.37 {\AA} for Kr 4$p$. 

The pseudopotentials are generated by the (semi)local approximations in DFT. This means that the HF-exchange energy between core and valence states are neglected. Yet we have found that the calculated energy bands obtained 
with the pseudopotentials generated in LDA and GGA, 
are essentially identical to each other, implying that the treatment of the exchange-correlation energy in generating pseudopotentials has minor effects.  

The partial core correction\cite{louie} is not included in our calculations. This is partly because magnetic properties are not considered in the present work. However, to assure the accuracy of structural properties, we regard some of core orbitals as valence orbitals and include them explicitly in the pseudopotential generation: 
Such orbitals included as valence states are $2s$ and $2p$ orbitals of Na and $2s$ and $2p$ orbitals of Mg. 

Appropriate choice of cutoff energies $E_{\rm cut}$ in the plane-wave-basis set, which is related to hardness of the adopted norm-conserving pseudopotentials, is a principal ingredient to assure the accuracy of the results. We have examined convergence of structural properties and band gaps with respect to $E_{\rm cut}$ and reached the following well converged values with 
$E_{\rm cut}$ for each material: 25 Ryd for Si and Ge, 36 Ryd for Kr, 64 Ryd for BaTiO$_3$, and 100 Ryd for diamond, GaN, MgO, NaCl, LiCl, and LiF.  
The remaining important ingredient to assure our assessment of each hybrid functional is the sampling $\kvec$ points for the BZ integration. We have adopted the scheme by Monkhorst and Pack in which BZ is divided by equally spaced mesh. After careful examination, we have found that 4$\times$4$\times$4 sampling $\kvec$ points are enough to assure the accuracy of the total energies and energy bands in the ten materials. The results are confirmed by repeating the calculations with 6$\times$6$\times$6 sampling $\kvec$ points. 
%

\begin{figure*}
\begin{center}	
\includegraphics[width=0.7\textwidth]{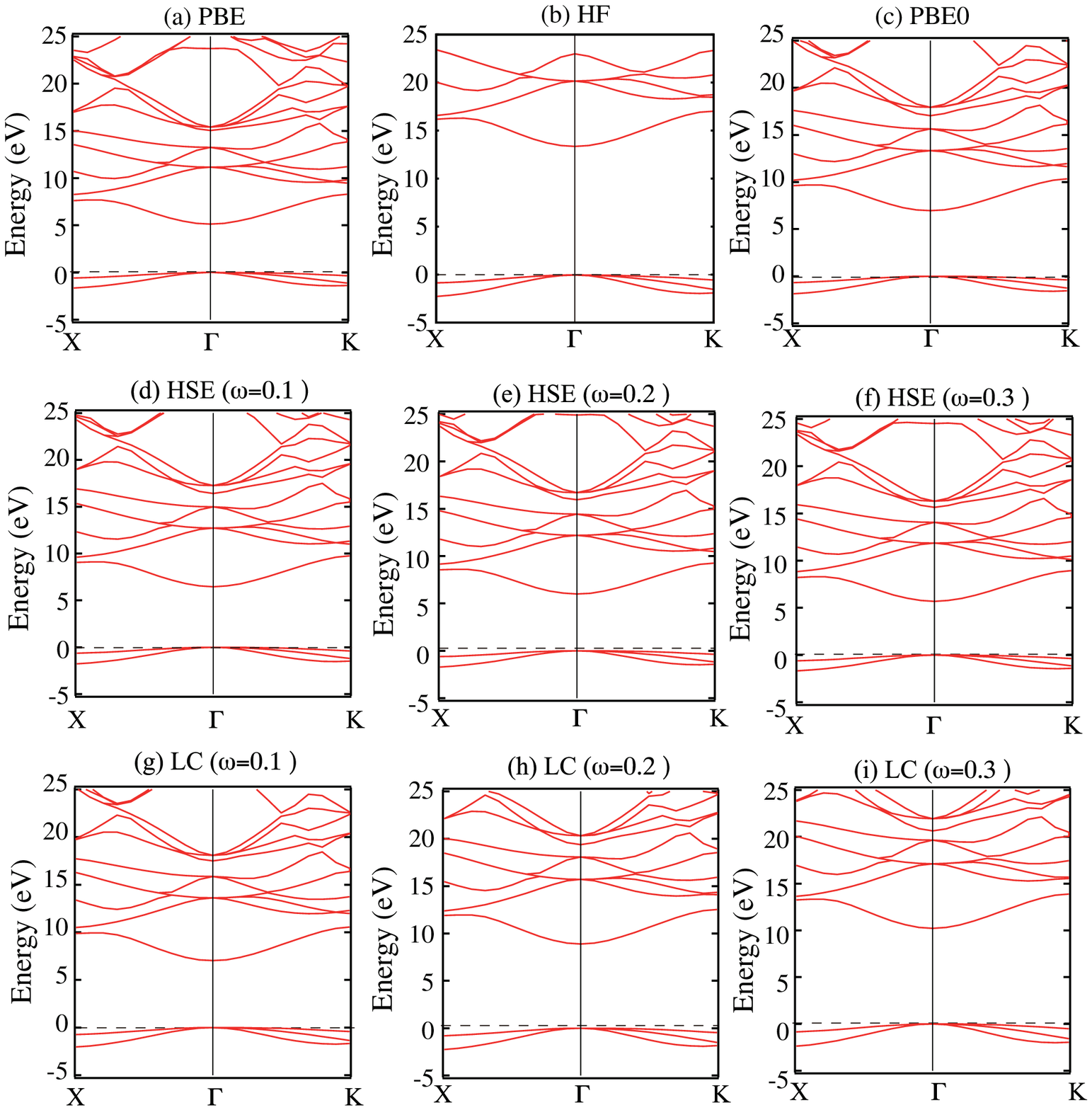}  
\end{center}
\caption{(Color online) Calculated energy bands of NaCl with PBE (a), HF (b), PBE0 (c), HSE (d)-(f), and LC (g)-(i) exchange-correlation functionals. The origin of the energy is set at the valence band top.}
\label{NaCl_band}
\end{figure*}

\section{Results and Discussion}\label{result}

\begin{table*}[hbtp]
\caption{Band gaps $\epsilon_{\rm gap}$ obtained from PBE, HF, PBE0, HSE, and LC calculations. The $\omega$ is a parameter which separates the long-range and short-range parts of the Coulomb interaction (see text). Experimental values are taken from Ref.~\onlinecite{ zahlenwerte} for Ge, Si, and C, Ref.~\onlinecite{heyd2005} for $\beta$-GaN, Ref.~\onlinecite{ref1} for BaTiO$_3$, Ref.~\onlinecite{adachi} for MgO, Ref.~\onlinecite{poole} for NaCl, Ref.~\onlinecite{ref2} for Kr, and Ref.~\onlinecite{rohlfing} for LiCl, and LiF. The calculated mean relative error (MRE) and the mean absolute relative error (MARE) with respect to experimental values are also shown in percent. Group I consists of materials having the experimental gap less than 7 eV, while Group II of materials with the gap more than 7 eV (see text).}
\label{bandgap}
\begin{center}
{\scriptsize 
\begin{tabular}{l@{\ \ \ }r@{\ \ \ }r@{\ \ \ }r@{\ \ \ \ \ \ }r@{\ \ \ }r@{\ \ \ }r@{\ \ \ }r@{\ \ \ \ \ \ }r@{\ \ \ }r@{\ \ \ }r@{\ \ \ }r@{\ \ \ \ \ \ }r}
\hline \hline
\multicolumn{13}{c}{$\epsilon_{\rm{gap}}$} 
\\ [2pt] \hline \\ [-4pt] 
 & PBE & HF & PBE0 &\multicolumn{4}{c}{HSE} & \multicolumn{4}{c}{LC} & Expt. 
\\ [2pt] \hline \\ [-4pt]  
 & & & & $\omega$=0.1 & $\omega$=0.2 & $\omega$=0.3 & $\omega$=0.4 
       & $\omega$=0.1 & $\omega$=0.2 & $\omega$=0.3 & $\omega$=0.4 & 
\\ \hline \\
Ge    &0&4.75&1.00&0.76 &0.54 &0.43 &0.27&1.05&2.05&2.66&3.69&0.74\\
Si    &0.61&6.03&1.72  &1.20 &0.94 &0.80&0.73&2.24&3.85&4.19&4.63&1.17\\
BaTiO$_3$&2.14&11.62   &4.21 &3.57 &3.12&2.83&2.73&4.59&6.26&7.45&8.01&3.2\\
$\beta$-GaN&2.06&8.84  &3.51 &3.02 &2.65&2.42&2.26&3.95&5.58&6.55&7.36&3.30\\
C     &4.01&12.44&5.87 &5.28 &4.88 &4.62&4.45&5.06&6.80&7.95&8.73&5.48\\
MgO   &4.95&14.21&6.99 &6.62 &6.14 &5.79&5.52&6.38&8.33&9.88&11.12&7.7\\
NaCl  &5.13&13.38&6.95 &6.46 &6.00 &5.69&5.49&7.03&8.90&10.22&11.12&8.5\\
LiCl  &6.33&14.85&8.60 &8.14 &7.67 &7.36&7.15&8.53&10.45&11.75&12.66&9.4\\
LiF &9.70&21.57&12.51&12.02&11.45&11.02&10.68&11.59&13.87&15.65&17.06&14.30\\ 
Kr    &7.09&15.22&9.14 &8.46 &7.98 &7.68 &7.48  & 9.85&11.75&13.00&13.75 &11.65
\\ [2pt] \hline \\ [-4pt] 
 \multicolumn{13}{c}{All solids} \\ [2pt] 
MARE (\%)&42.5&179.7&19.7&12.4&19.9&26.4&31.5&28.2&62.3&88.9&117.4\\
MRE (\%)&$-$42.5&179.7&5.7&$-$9.0&$-$19.9&$-$26.4&$-$31.5&11.1&61.7&88.9&117.4\\  [2pt] \hline \\ [-4pt]
 \multicolumn{13}{c}{Group I (Ge, Si, BaTiO$_3$, GaN, C)} \\ [2pt] 
 MARE (\%)&49.1&303.1&25.4&5.8&16.0&25.5&33.2&40.8&119.0&158.8&205.4\\
MRE (\%)&$-$49.1&303.1&25.4&0.9&$-$16.0&$-$25.5&$-$33.2&37.8&119.0&158.8&205.4\\  [2pt] \hline \\ [-4pt]
\multicolumn{13}{c}{Group II (MgO, NaCl, LiCl, LiF, Kr)} \\ [2pt] 
MARE (\%)&35.9&56.3&14.0&19.0&23.9&27.3&29.8&15.6&5.6&18.9&29.4\\
MAE (\%)&$-$35.9&56.3&$-$14.0&$-$19.0&$-$23.9&$-$27.3&$-$29.8&$-$15.6&4.4&18.9&29.4\\ \hline \hline
\end{tabular}
} 
\end{center}
\end{table*}

We have performed total-energy electronic-structure calculations using PBE, HF, PBE0, HSE, LC exchange-correlation functionals for ten materials, including covalent semiconductors, ionic insulators, dielectric compounds, and a rare-gas solid: i.e., Si, Ge, GaN, diamond, MgO, NaCl, LiCl, LiF, BaTiO$_3$, and Kr. The calculated results elucidate capability and limitation of each functional in discussing electronic and structural properties of these prototype materials. We first present calculated electron states of these materials and compare them with experimental results 
in the subsection \ref{ele_state}. Then we present the results of structural optimization in the subsection \ref{structure}.

\subsection{Energy bands and gaps}\label{ele_state} 
\begin{table*}[hbtp]
\caption{Valence bandwidth $W$ obtained from PBE, HF, PBE0, HSE, and LC calculations. The $\omega$ is a parameter which separates the long-range and short-range parts of the Coulomb interaction (see text). Experimental values are taken from Refs.~\onlinecite{GW1, wachs} for Ge, Refs. \onlinecite{GW1, zahlenwerte} for Si,
Refs. \onlinecite{GW1,mcfeely,himpsel} for C.} 
\label{bandwidth}
\begin{center}
{\scriptsize 
\begin{tabular}{l@{\ \ \ }r@{\ \ \ }r@{\ \ \ }r
@{\ \ \ \ \ \ \ \ }r@{\ \ \ }r@{\ \ \ }r
@{\ \ \ \ \ \ \ \ }r@{\ \ \ }r@{\ \ \ }r
@{\ \ \ \ \ \ }r}
\hline \hline \\ [-2pt] 
\multicolumn{11}{c}{$W$}
\\ [2pt] \hline \\ [-4pt] 
 &PBE&HF&PBE0& \multicolumn{3}{c}{HSE}&\multicolumn{3}{c}{LC}& Expt. 
\\ [2pt] \hline \\ [-4pt]  
 &   &  &    & $\omega$=0.1 & $\omega$=0.2 & $\omega$=0.3 
             & $\omega$=0.1 & $\omega$=0.2 & $\omega$=0.3 & 
\\ [2pt] \hline \\ [-4pt] 
Ge&12.85&18.22&14.50&13.80&13.47&13.11&14.23&15.54&16.36&12.9$\pm$0.2\\
Si&11.95&16.90&13.37&13.28&12.99&12.65&12.19&13.45&14.52&12.5$\pm$0.6\\
BaTiO$_3$  &4.53&6.37&5.08&5.03&4.95&4.85&4.85&5.34&5.72 & \\
$\beta$-GaN&6.73&8.27&7.20&7.11&7.02&6.93&7.11&7.44&7.91 & \\
C&21.65&30.09&23.64&23.51&23.28&22.96&24.15&25.00&26.19 
&24.2$\pm$1,\ \ 21$\pm$1 \\ 
MgO &4.43&6.16&4.97 &4.80 &4.70 &4.60&5.21&5.56&5.92& \\
NaCl&1.65&2.24&1.84 &1.77 &1.71 &1.67&2.04&2.25&2.38& \\
LiCl&2.82&4.33&3.39 &3.16 &3.16 &3.09&3.59&3.91&4.17& \\
Kr  &1.50&1.93&1.61 &1.59 &1.55 &1.52&1.57&1.74&1.85&\\
LiF &2.83 &3.66&3.09 &2.96 &2.92 &2.87 &2.94&3.09&3.27& 
\\ \hline \hline
\end{tabular}
} 
\end{center}
\end{table*}

Figure \ref{Si_band} shows calculated band structures of Si with five exchange-correlation functionals. For the HSE and LC functionals, the band structures with different choices of $\omega$ [0.1, 0.2, and 0.3 ($a_B^{-1}$) ] in Eq.~(\ref{split}) are shown. The overall features of the band structures obtained by the five functionals are similar to each other. Yet the bandwidths and the fundamental gaps are different quantitatively. When the ratio of the HF exchange to the total-exchange functional is large, the resulting bandwidth and gap become large; these quantities become larger in the order of PBE, PBE0, HSE, LC, and HF. Notice that the band gaps obtained by the LC functional are always larger than those by the HSE functional, indicating that the correction to the long-range part of the exchange potential tends to make the band gap large. 

We also show the calculated energy bands of the dielectric compound BaTiO$_{3}$ and the ionic insulator NaCl in Figs.~\ref{BaTiO3_band} and \ref{NaCl_band}, respectively. We find the general tendency similar to that in Si: i.e., the overall features of the energy bands are insensitive to the difference in the functionals; the LC functional provides larger energy gaps compared with the HSE functional. 

TABLE~\ref{bandgap} and \ref{bandwidth} summarize the calculated bandwidths and band gaps for the ten materials. For assessment of validity of each functional, it is convenient to introduce two quantities, the mean relative error (MRE) 
which is the mean of the calculated value minus the experimental value over the ten materials, and the mean absolute relative error (MARE) 
which is the mean of the absolute value of the difference between the calculated and the experimental values over the ten materials. The MRE is a measure of under- or over-estimates of the experimental values since each functional predicts either smaller or larger values than the experimental values for most of the ten materials. On the other hand, MARE is a measure of closeness between the calculated and experimental values. 
In the discussion below, we categorize the ten materials into Group I and Group II: Group I, i.e., diamond, GaN, BaTiO$_3$, Si and Ge has covalent characters in which the band gaps are less than 7 eV; Group II, i.e., LiF, Kr, NaCl, LiCl and MgO, consists of ionic solids and a rare-gas solid in which the experimental band gaps are larger than 7 eV.   

The calculated band gap by the PBE functional for each material is substantially smaller than the corresponding experimental value, as is reported in literature. The calculated MRE for the ten materials is $-$42.5 \%. On the other hand, HFA largely overestimates the band gaps for all the ten materials: MRE is 179.7 \%. The PBE0 functional provides better values. The calculated MRE by the PBE0 for the materials in Group I is 25.4 \%, whereas it gives $-$14.0 \% for the materials in Group II. 

The HSE and LC functional also provide better agreement of the band gap with the experiments than the PBE does. Further when we choose optimum values of $\omega$, the HSE and LC results show nice agreement with the experimental values. In the HSE functional, a general trend is the decrease in the band gap with increasing $\omega$. We have found that the value $\omega$ = 0.1 $a_B^{-1}$ produces the band gaps close to the experimental values within MRE = 0.9 \% for the materials in Group I. It produces worse values for the materials in Group II. The calculated MRE is $-$19.0 \% for the materials in Group II, but still HSE being the better approximation than PBE. 

The LC functional provides good agreement with the experimental values for the materials in Group II: When we choose the optimum value of $\omega = 0.2 a_B^{-1}$, the calculated MRE is nicely small, 4.4 \% for the materials in Group II.  Yet the LC provides worse values for the materials in Group I, showing its limitation as an universally valid approximation. 

\begin{figure*}
\begin{center}
\includegraphics[width=0.6\textwidth]{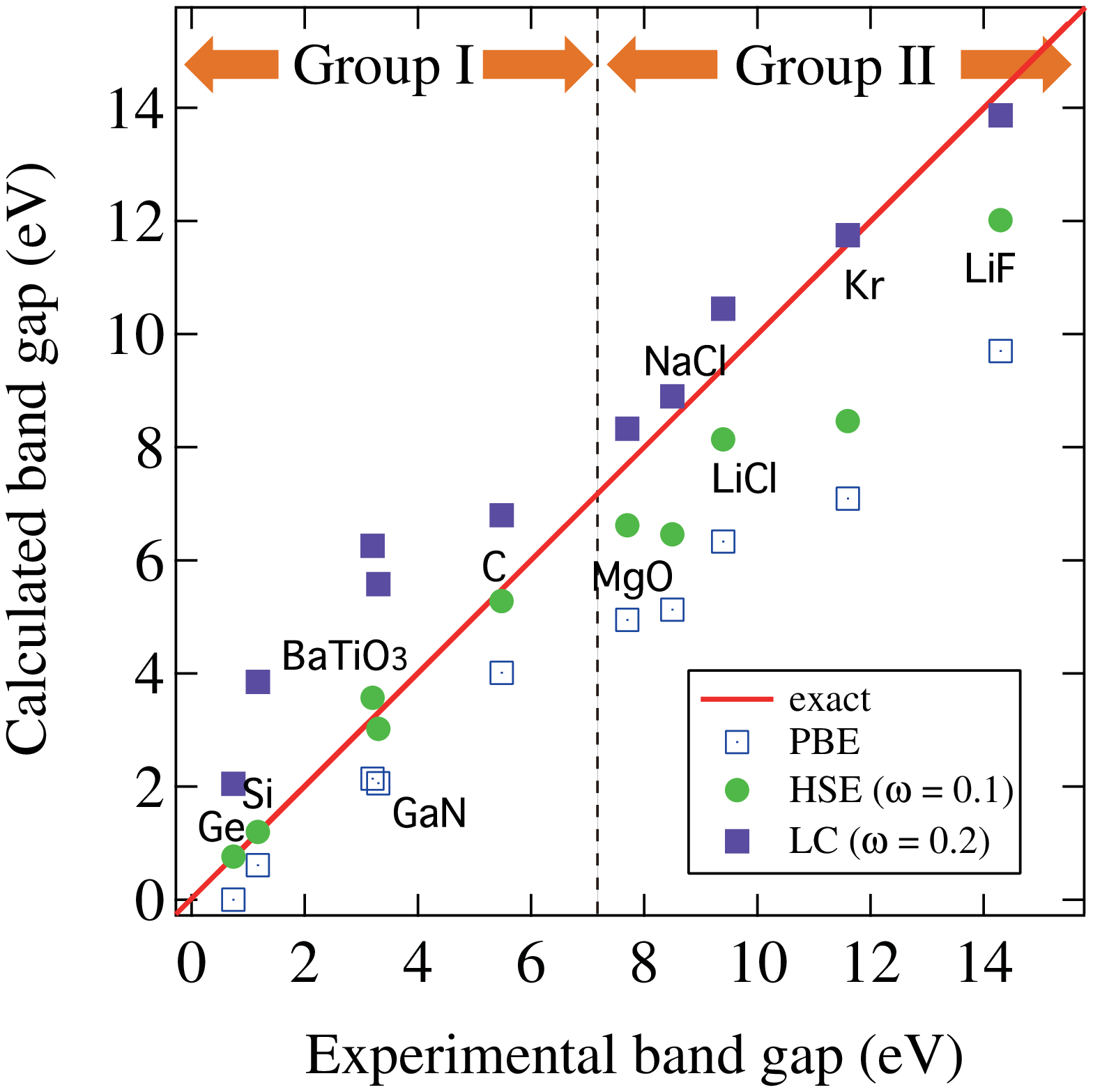}  
\end{center}
\caption{(Color online) Calculated band gaps obtained from different exchange-correlation functionals: PBE (blank squares), HSE with $\omega$=0.1 $a_{B}^{-1}$ (green dots), and LC with $\omega$=0.2 $a_{B}^{-1}$ (purple squares), plotted against experimental band gaps. Group I consists of materials having the experimental gap less than 7 eV, while Group II is composed of materials with the gap more than 7 eV.}
\label{theory_vs_expt}
\end{figure*}

Figure~\ref{theory_vs_expt} is a summary of our calculated band gaps by the PBE, HSE and LC functionals. For HSE and LC, we show the calculated results with the optimum $\omega$ values: $\omega$ = 0.1 $a_B^{-1}$ for HSE and $\omega$ = 0.2 $a_B^{-1}$ for LC. It is clearly shown that the calculated band gaps 
by hybrid functionals, HSE and LC, are in better agreement with the experimental values than the PBE (GGA) approximation, indicating the promising possibility of the hybrid functionals. The degree of the agreement is close to the Green's-function-based GW approximation in which there are several ambiguities in theoretical treatments.\cite{GW1,GW2,GWissue}  
However, this figure also shows the limitation; HSE is reasonably good for only the Group I materials, whereas LC is good for only the Group II materials.  

Our finding here is that HSE is a good approximation for relatively small-gap materials and that LC is a good approximation for the relatively large-gap materials. Screening of the Coulomb interaction depends on the materials. Hence the best treatment of the short-range and long-range parts of the Coulomb interactions should change material by material. What we have found here is natural in the sense. However, another point we have shown here is that the appropriate choice of $\omega$ with each exchange-correlation functional provides reasonable agreement with the calculated band gaps for rather wide range of materials; HSE with $\omega$ $\sim$ 0.1 $a_B^{-1}$ for the materials with their band gaps smaller than 7 eV 
and LC with $\omega$ $\sim$ 0.2 $a_B^{-1}$ for the materials with their band gaps larger than 7 eV. This information certainly provides a practical recipe to obtain reliable band gaps for various materials.  

The calculated bandwidths by the PBE functional are in good agreements with the experimental values available (TABLE~\ref{bandwidth}). Other hybrid functionals also provide reasonable agreement, in particular, with appropriate choices of the $\omega$ parameter for the HSE and LC functionals. On the other hand, HFA shows substantial overestimate for each material.  

\subsection{Structural properties}\label{structure}

\begin{table*}
\caption{Lattice constants $a_0$ (\AA) obtained from PBE, HF, PBE0, HSE, and LC calculations. The $\omega$ parameter separates the short-range and long-range parts of the Coulomb interaction (see text). 
Experimental values are taken from Ref.~\onlinecite{heyd2005} for Ge, Si, $\beta$-GaN, C, and MgO, Ref.~\onlinecite{piskunob,hellwege} for BaTiO$_3$, Ref.~\onlinecite{heyd2004} for NaCl, LiCl, and LiF, and Ref.~\onlinecite{varotsos} for Kr.
The calculated mean relative error (MRE) and the mean absolute relative error (MARE) with respect to experimental values are also shown in per cent. }
\label{lattice}
\begin{center}
{\scriptsize 
\begin{tabular}{l@{\ \ \ }r@{\ \ \ }r@{\ \ \ }r@{\ \ \ \ \ \ }r@{\ \ \ }r@{\ \ \ }r@{\ \ \ }r@{\ \ \ \ \ \ }r@{\ \ \ }r@{\ \ \ }r@{\ \ \ }r@{\ \ \ \ \ \ }r}
\hline \hline \\ [-4pt] 
\multicolumn{13}{c}{$a_0$} 
\\ [2pt] \hline \\ [-4pt] 
 & PBE & HF & PBE0 & \multicolumn{4}{c}{HSE}&\multicolumn{4}{c}{LC}& Expt. 
\\ [2pt] \hline \\ [-4pt]  
 & & & & $\omega$=0.1 & $\omega$=0.2 & $\omega$=0.3 & $\omega$=0.4  
       & $\omega$=0.1 & $\omega$=0.2 & $\omega$=0.3 & $\omega$=0.4 & 
\\ [2pt] \hline \\ [-4pt] 
Ge&5.589&5.574&5.615&5.546&5.556&5.566&5.571&5.579&5.534&5.508&5.490&5.652\\
Si&5.463&5.387&5.431&5.435&5.441&5.446&5.452&5.459&5.434&5.416&5.403&5.430\\
BaTiO$_3$&4.139&4.048&4.068&4.102&4.106&4.111&4.116&4.137&4.120&4.102&4.088&4.000\\
$\beta$-GaN&4.539&4.380&4.434&4.455&4.457&4.441&4.461&4.502&4.511&4.491&4.452&4.520\\
C&3.563&3.485&3.506&3.522&3.540&3.543&3.547&3.560&3.556&3.543&3.531&3.567\\
MgO&4.202&4.198&4.198&4.117&4.136&4.141&4.146&4.178&4.175&4.157&4.138&4.207\\
NaCl&5.541&5.416&5.444&5.500&5.505&5.509&5.514&5.525&5.503&5.487&5.472&5.595\\
LiCl&5.175&5.107&5.133&5.145&5.147&5.155&5.158&5.165&5.140&5.117&5.115&5.106\\
Kr    &10.86&10.06&10.74&10.79&10.81&10.86&10.86&10.69&10.49&10.05&10.05&9.94\\
LiF&4.115&4.006&4.030&4.041&4.071&4.073&4.076&4.103&4.098&4.091&4.080&4.010 
\\ [2pt] \hline \\ [-4pt] 
\multicolumn{13}{c}{All solids} \\ [2pt] 
MARE (\%)&1.20&1.37&1.10&1.40&1.37&1.41&1.35&1.25&1.21&1.34&1.54\\
MRE  (\%)&0.68&$-$1.10&$-$0.49&$-$0.47&$-$0.22&$-$0.16&$-$0.03&0.40&0.10&$-$0.27&$-$0.62\\ [2pt] \hline \\ [-4pt]  
\multicolumn{13}{c}{Group I (Ge, Si, GaN, BaTiO$_3$, C)} \\ [2pt] 
MARE (\%)&1.15&1.75&1.20&1.44&1.34&1.40&1.32&1.17&1.13&1.33&1.62\\
MRE  (\%)&0.66&$-$1.27&$-$0.51&$-$0.39&$-$0.20&$-$0.17&0.00&0.41&0.10&$-$0.31&$-$0.74
\\ [2pt] \hline \\ [-4pt]  
\multicolumn{13}{c}{Group II (MgO, NaCl, LiCl, LiF)}\\ [2pt] 
MARE (\%)&1.26&0.88&0.99&1.34&1.41&1.41&1.39&1.35&1.32&1.34&1.44\\
MRE  (\%)&0.72&$-$0.87&$-$0.47&$-$0.58&$-$0.24&$-$0.14&$-$0.06&0.38&0.11&$-$0.22&$-$0.48
\\ [2pt] \hline \hline
\end{tabular}
} 
\end{center}
\end{table*}

We next examine performance of the hybrid functionals in describing structural properties. TABLE~\ref{lattice} and \ref{BM} show calculated lattice constants $a_0$ and bulk moduli $B_0$, respectively, of the ten materials. The $a_0$ and $B_0$ values are determined 
by fitting parameters in the Murnaghan equation of state to the calculated total energy as a function of the volume. 
In MARE and MRE presented in TABLE~\ref{lattice}, the calculated value for solid Kr is not included. As shown in this TABLE, 
the calculated lattice constant of Kr is substantially larger than the experimental value: It is overestimated by 1.1-9.2 \% in GGA and in 
the hybrid approximations, and by 1.2 \% in HFA. 
This is because van der Waals interaction which is unable to be treated in the 
approximations examined in this paper plays an essential role in solid Kr. 
We thus exclude the value for Kr from the statistical assessment. 
The results obtained by PBE reasonably agree with the experimental values (MRE = 0.68 \% and MARE = 1.20 \% for $a_0$ and MRE = $-$7.1 \% and MARE = 8.3 \% for $B_0$). The accuracy of HFA is slightly inferior to that of PBE (MRE = $-$1.10 \% and MARE = 1.37 \% for $a_0$ and MRE = 13.7 \% and MARE = 13.7 \% for $B_0$). The PBE0 functional shows better accuracy with the MRE = $-$0.49 \% and MARE = 1.10 \% for $a_0$, and MRE = $-$1.7 \% and MARE = 5.1 \% for $B_0$. 

The calculated lattice constants using the HSE and LC functionals are insensitive to the choice of $\omega$: The difference obtained from different $\omega$ values is within 1 \%. The difference in the calculated bulk modulus from different $\omega$ values are not so small as in the lattice constant partly because the fitting by the Murnaghan equation is incomplete. For the HSE functional, the value $\omega$ = 0.1 $a_B^{-1}$ produces best agreement with the experiments: MRE = $-$0.47 \% and MARE = 1.40 \% for $a_0$, and MRE = 1.7 \% and MARE = 3.6 \% for $B_0$. For the LC functional, we have found that $\omega$ = 0.2 $a_B^{-1}$ produces best results: MRE = 0.10 \% and MARE = 1.21 \% for $a_0$, and MRE = 2.0 \% and MARE = 5.5 \% for $B_0$. These optimum values for $\omega$ for the HSE and LC functionals are identical to the optimum values determined from the calculated MRE and MARE for band gaps, corroborating the appropriate choice of $\omega$.

\begin{table*}
\caption{Bulk moduli $B_0$ (GPa) obtained from PBE, HF,PBE0, HSE, and LC calculations. 
Experimental values are taken from Ref.~\onlinecite{jivani} for Ge, Ref.~\onlinecite{heyd2004} for Si, $\beta$-GaN, C, MgO, NaCl, LiCl, and LiF, Ref.~\onlinecite{piskunob,hellwege} for BaTiO$_3$, and Ref.~\onlinecite{varotsos} for Kr.
 The calculated mean relative error (MRE) and the mean absolute relative error (MARE) with respect to experimental values are also shown in per cent. }
\label{BM}
\begin{center}
{\scriptsize 
\begin{tabular}{l@{\ \ \ }r@{\ \ \ }r@{\ \ \ }r@{\ \ \ \ \ \ }r@{\ \ \ }r@{\ \ \ }r@{\ \ \ }r@{\ \ \ \ \ \ }r@{\ \ \ }r@{\ \ \ }r@{\ \ \ }r@{\ \ \ \ \ \ }r}
\hline \hline \\ [-4pt] 
\multicolumn{13}{c}{$B_0$} 
\\ [2pt] \hline \\ [-4pt] 
& PBE& HF& PBE0& \multicolumn{4}{c}{HSE}&\multicolumn{4}{c}{LC}& Expt.  
\\ [2pt] \hline \\ [-4pt] 
 & & & & $\omega$=0.1 & $\omega$=0.2 & $\omega$=0.3 & $\omega$=0.4 
       & $\omega$=0.1 & $\omega$=0.2 & $\omega$=0.3 & $\omega$=0.4 &  
\\ [2pt] \hline \\ [-4pt] 
Ge&69.0&76.0&68.8&76.3&75.8&73.3&72.7&71.8&85.3&89.0&95.8&75.8\\
Si&91.1&116.3&98.1&95.9&93.8&92.0&90.4&90.4&98.0&104.3&109.0&99.2\\
BaTiO$_3$&146.0&184.0&167.0&164.0&162.0&159.0&156.0&150.0&167.0&166.0&175.0&162.0\\ 
$\beta$-GaN&173.0&274.6&220.3&215.5&213.3&203.5&189.1&192.2&195.0&218.1&249.3&210.0 \\ 
C   &456.0&534.0&488.0&485.0&483.0&478.0&473.0&469.0&472.0&490.0&508.0&443.0\\ 
MgO &152.0&181.0&169.0&166.0&164.0&162.0&160.0&152.0&154.0&161.0&169.0&165.0\\ 
NaCl&27.3&32.7&29.3&29.0&28.9&28.7&28.4&28.6&29.1&29.9&30.9&26.6\\ 
LiCl&30.7&37.3&33.6&33.6&33.5&33.3&33.1&34.2&36.3&35.1&36.0&35.4\\ 
Kr  &2.4   &6.4  &3.8  &3.8   &3.5  &3.1  &2.7   &3.2  &3.5  &4.7  &5.1  &3.4\\
LiF &67.2&72.0&69.9&69.8&69.8&69.6&69.2&73.1&69.3&67.9&72.0&69.8 
\\ [2pt] \hline \\ [-4pt] 
\multicolumn{13}{c}{All solids} \\ [2pt] 
MARE (\%)&8.3&13.7&5.1&3.6&3.4&4.4&5.6&6.6&5.5&6.4&11.2\\
MRE  (\%)&$-$7.1&13.7&1.7&1.7&0.9&$-$0.9&$-$2.6&$-$2.6&2.0&5.1&11.2
\\ [2pt] \hline \\ [-4pt] 
\multicolumn{13}{c}{Group I (Ge, Si, GaN, BaTiO$_3$, C)} \\ [2pt] 
MARE (\%)&9.5&16.5&5.7&3.5&3.2&4.7&6.7&7.2&6.1&7.9&15.5\\
MRE  (\%)&$-$8.3&16.5&1.6&2.1&1.0&$-$1.5&$-$4.0&$-$4.8&2.8&7.9&15.5
\\ [2pt] \hline \\ [-4pt] 
\multicolumn{13}{c}{Group II (MgO, NaCl, LiCl, LiF)}\\ [2pt] 
MARE (\%)&6.9&10.3&4.5&3.7&3.7&4.0&4.3&5.9&4.8&4.6&5.9\\
MRE  (\%)&$-$5.6&10.3&1.9&1.1&0.7&0.0&$-$0.9&0.2&1.1&1.6&5.9
\\ [2pt] \hline \hline
\end{tabular}
}
\end{center}
\end{table*}

\section{Conclusion}\label{conclusion}

We have studied validity of hybrid exchange-correlation functionals in density functional theory by implementing three hybrid functionals in a well-established plane-wave-basis-set code named TAPP and by calculating structural properties and electron states of representative ten materials where the experimental energy gaps range from 0.67 eV to 14.20 eV. The three hybrid exchange-correlation functionals examined in this paper are PBE0 proposed by Perdew, Burke, and Ernzerhoff, HSE proposed by Heyd, Scuseria, and Ernzerhoff, and LC originally proposed by Savin and by Hirao and his collaborators. For comparison, Results from the generalized gradient approximation, i.e., PBE, and HFA have been presented. The ten materials we have examined are Ge, Si, GaN, BaTiO$_3$, diamond, MgO, LiCl, NaCl, Kr, and LiF which are representatives of covalent, ionic, and rare-gas solids. 

We have found that the structural properties such as lattice constants are already well reproduced by the PBE functional and also by HFA and that the hybrid functionals show better agreement with the experimental values. We have determined appropriate values of $\omega$ in the separation of the short-range and long-range parts in the Coulomb interaction: The optimum value is $\omega$ = 0.1 $a_B^{-1}$ for the HSE functional and $\omega$ = 0.2 $a_B^{-1}$ for the LC functional. By choosing the appropriate value of $\omega$ in the HSE and LC functionals, we have achieved better agreement in the lattice constants and further substantial improvement in description of elastic constants such as bulk moduli for the ten materials. 

Dramatic success of the hybrid functionals are observed in the calculated band gaps. We have found that the calculated band gaps by the LC functional for the wide band-gap materials satisfactorily agree with the experimental values with mean relative error (MRE) of 3.0 \%, whereas the band gaps by the HSE functional for the small band-gap materials agree well with the experimental values with MRE = 0.7 \%. This good description of the band gaps is unprecedented in density functional theory where LDA and GGA produce the value of MRE $\sim$ 40-50 \%, and is comparable with or better than what the GW approximation produces. The $\omega$ value leading to the best agreement with the experiments is 
0.1 $a_B^{-1}$ for the HSE functional and 
0.2 $a_B^{-1}$ for the LC functional. These $\omega$ values are identical to the optimum values determined from the examination of the structural properties. The calculated valence-band widths by the hybrid functionals also agree satisfactorily with the experimental values. 

It is now established that the HSE and LC functionals with appropriate choice of the $\omega$ parameter are useful to describe structural and electronic properties of various materials. Rigorous justification of the choice of the form of the hybrid functionals along with a guiding principle of choice of $\omega$ would offer further developments in the first-principles calculations.

\acknowledgments

The work is partly supported by a grant-in-aid project from MEXT, Japan, ``Scientific Research on Innovative Areas: Materials Design through Computics - Complex Correlation and Non-equilibrium Dynamics -" under the contract number 22104005. Computations were done at Supercomputer Center, Institute for Solid State Physics, University of Tokyo, and at Research Center for Computational Science, National Institutes of Natural Sciences.

\end{document}